\documentclass[epj,referee]{svjour}

\begin{document}

\title{Wronskian perturbation theory}

\author{Paolo Amore\inst{1} \and Francisco M. Fern\'andez\inst{2}}

\institute{Facultad de Ciencias, Universidad de Colima, Bernal
D\'{i}az del Castillo 340, Colima, Colima, Mexico
\email{paolo@ucol.mx} \and INIFTA (Conicet, UNLP), Blvd. 113 y 64
S/N, Sucursal 4, Casilla de Correo 16, 1900 La Plata, Argentina
\email{fernande@quimica.unlp.edu.ar} \mail{F.M.F.}}
\abstract{ We develop a perturbation method that generalizes an
approach proposed recently to treat velocity--dependent
quantum--mechanical models. In order to test present approach we
apply it to some simple trivial and nontrivial examples.
     }
\maketitle
\section{Introduction}

In a recent paper Jaghoub~\cite{J06} developed a perturbation
theory for velocity--dependent quantum--mechanical models. Such
potentials are useful, for example, for the study of pion--nucleon
scattering and in models of particles with coordinate--dependent
masses~\cite{J06} (and references therein). According to the
author, one of the advantages of the method is that its main
equations depend only on the eigenfunction of the chosen
unperturbed state. More precisely, the method does not need
neither the whole unperturbed energy spectrum nor the basis set of
unperturbed eigenfunctions as in the formulation of the
Rayleigh--Schr\"{o}dinger perturbation theory in terms of sums
over intermediate states. Jaghoub chose rather too simple examples
in order to test his method~\cite{J06}.

Similar perturbation approaches are known since long ago, but they
apply mainly to local (coordinate--dependent)
potentials~\cite{BO78}. Although one can in principle adapt some
of the well--known perturbation methods~\cite{F01} to the
treatment of velocity--dependent potentials, here we proceed in a
different way.

The purpose of this paper is to generalize the method proposed by
Jaghoub~\cite{J06} and derive a perturbation algorithm for the
treatment of a wider variety of problems. In Sec.~\ref{sec:Method}
we develop our version of the method in a quite general way. In
Sec. \ref{sec:solvable} we apply the perturbation approach to an
exactly solvable example which enables us to compare the results
of present procedure with the expansion of the exact
eigenfunctions an eigenvalues. In Sec.~\ref{sec:nonsolvable} we
discuss a
partially solvable example treated by Jaghoub~\cite{J06}. Finally, in Sec.~%
\ref{sec:conclusions} we summarize our main results and discuss
other potential applications of present method.

\section{Method \label{sec:Method}}

Suppose that we want to solve the differential equation
\begin{equation}
y^{\prime \prime }(x)=F(y,x)  \label{eq:y"}
\end{equation}
where the prime denotes differentiation with respect to $x$ and
$F(y,x)$ an arbitrary linear differential operation on $y(x)$. In
order to apply perturbation theory we choose a closely related,
solvable problem
\begin{equation}
y_{0}^{\prime \prime }(x)=F_{0}(y_{0},x).  \label{eq:y0"}
\end{equation}
If we multiply Eq.~(\ref{eq:y"}) by $y_{0}$ and subtract
Eq.~(\ref{eq:y0"}) multiplied by $y$ we obtain
\begin{equation}
\frac{d}{dx}W(y,y_{0})=Fy_{0}-F_{0}y  \label{eq:dW/dx}
\end{equation}
where $W(y,y_{0})$ stands for the Wronskian
\begin{equation}
W(y,y_{0})=y^{\prime }y_{0}-y_{0}^{\prime }y=y_{0}^{2}\left( \frac{y}{y_{0}}%
\right) ^{\prime }.  \label{eq:W}
\end{equation}

On integrating Eq.~(\ref{eq:dW/dx}) twice we obtain
\begin{eqnarray}
y(x) &=&C_{2}y_{0}(x)+C_{1}y_{0}(x)\int_{\beta }^{x}\frac{dx^{\prime }}{%
y_{0}(x^{\prime })^{2}}+  \nonumber \\
&&y_{0}(x)\int_{\beta }^{x}\frac{dx^{\prime }}{y_{0}(x^{\prime })^{2}}%
\int_{\alpha }^{x^{\prime }}\left( Fy_{0}-F_{0}y\right) (x^{\prime
\prime })dx^{\prime \prime }  \label{eq:y(x)}
\end{eqnarray}
where the integration constants $C_{1}$ and $C_{2}$ and the
integration limits $\alpha $ and $\beta $ enable one to
accommodate the boundary conditions of the problem and the
normalization of the solution.

Notice that the function
\begin{equation}
u(x)=y_{0}(x)\int_{\beta }^{x}\frac{dx^{\prime }}{y_{0}(x^{\prime
})^{2}} \label{eq:u(x)}
\end{equation}
satisfies $W(u,y_{0})=1$, so that $u(x)$ does not vanish and is
finite at
the zeroes of $y_{0}(x)$. If the operation $F_{0}$ is simply of the form $%
F_{0}(y_{0},x)=f_{0}(x)y_{0}(x)$, then $y_{0}(x)$ and $u(x)$ are
two linearly independent solutions of Eq.~(\ref{eq:y0"}), and
$u(x)$ is called ghost state~\cite{ACCCY91} (and references
therein). A complex linear combination of $y_{0}(x)$ and $u(x)$
proved suitable for the construction of a logarithmic perturbation
method for excited states~\cite{ACCCY91}.

In order to apply perturbation theory we introduce a perturbation parameter $%
\lambda $ into $F$ and expand
\begin{equation}
F=\sum_{j=0}^{\infty }F_{j}\lambda ^{j}  \label{eq:F_series}
\end{equation}
and
\begin{equation}
y(x)=\sum_{j=0}^{\infty }y_{j}(x)\lambda ^{j}.
\label{eq:y_series}
\end{equation}
We simply introduce the series~(\ref{eq:F_series}) and
(\ref{eq:y_series})
into Eq.~(\ref{eq:y(x)}) and obtain $y_{j}(x)$ in terms of $y_{k}(x)$, $%
k=0,1,\ldots ,j-1$, and $E_{k}$, $k=1,2,\ldots ,j$. The boundary
conditions determine the appropriate value of $E_{j}$ that is the
only unknown in the expression of $y_{j}(x)$. The procedure will
be made more explicit in the examples below.

Present approach applies to bound and unbound states; in this
paper we concentrate on the former ones. It is customary to choose
a convenient normalization for bound states. Here we keep our
equations as simple as possible and add a normalization factor
$N(\lambda )=N_{0}+N_{1}\lambda +\ldots $ at the end of the
calculation if necessary. We may, for example, require that
$(N_{0}+N_{1}\lambda +\ldots )(y_{0}+y_{1}\lambda +\ldots )$ be
normalized to unity up to a given perturbation order. Assuming
that $y_{0}(x) $ is normalized to unity we easily prove that
\begin{eqnarray}
N(\lambda )=1&&-\lambda \int y_{0}y_{1}dx+\frac{\lambda
^{2}}{2}\left[ 3\left( \int y_{0}y_{1}dx\right) ^{2}\right.\\
&&\left. -2\int y_{0}y_{2}dx-\int y_{1}^{2}dx\right] +\ldots ,
\label{eq:N(lambda)}
\end{eqnarray}
where the integrals extend over the whole physical coordinate
range. Obviously, the eigenvalues are independent of the
normalization constant.

It is worth noticing that we have not made explicit use of the
assumed linearity of $F(y,x)$ in order to derive Eq.
(\ref{eq:y(x)}). Consequently,
that equation applies to arbitrary nonlinear differential operators $F$ and $%
F_{0}$. However, the eigenvalues of nonlinear equations do depend
on the chosen normalization condition and we have to take this
fact explicitly into account in order to solve them.

\section{Solvable example \label{sec:solvable}}

In order to test the general equations developed above we choose
the eigenvalue problem
\begin{eqnarray}
y^{\prime \prime }(x) &=&\lambda y^{\prime }(x)-Ey(x),  \nonumber \\
y(0) &=&y(1)=0  \label{eq:model1}
\end{eqnarray}
with solutions
\begin{eqnarray}
y(x) &=&Ae^{\lambda x/2}\sin (n\pi x)  \nonumber \\
E &=&n^{2}\pi ^{2}+\frac{\lambda ^{2}}{4},\;n=1,2,\ldots
\label{eq:model1_sols}
\end{eqnarray}
Notice that the eigenvalue equation~(\ref{eq:model1}) is not
Hermitian but it supports real eigenvalues for real $\lambda $.
For the time being the exact form of the normalization factor $A$
is unnecessary for present perturbation calculation and we can add
it easily at the end of the process as discussed above. In order
to simplify the presentation and discussion of the results we
arbitrarily choose the normalization factor for the unperturbed
solution: $A=A(\lambda =0)=\sqrt{2}$.

Straightforward expansion of the exact
results~(\ref{eq:model1_sols}) in a Taylor series about $\lambda
=0$ yields
\begin{eqnarray}
y_{j}(x) &=&\frac{x^{j}}{j!2^{j}}y_{0}(x)  \nonumber \\
E_{j}(n) &=&n^{2}\pi ^{2}\delta _{j0}+\frac{1}{4}\delta _{j2}.
\label{eq:model1_yj_Ej}
\end{eqnarray}
In order to apply present perturbation theory to this simple test
model we
make the obvious choice $F=\lambda y^{\prime }(x)-Ey(x)$ and $%
F_{0}=-E_{0}y_{0}(x)$, where $E_{0}(n)=n^{2}\pi ^{2}$ and $y_{0}(n,x)=\sqrt{2%
}\sin (n\pi x)$. Since $u(x)$ does not vanish at the end points we choose $%
C_{1}=0$. The other integration constant $C_{2}$ is important to
force a normalization condition at each order of the perturbation
algorithm. However, as in this first step we are not interested in
the normalization of the solution we simply set $C_{2}=0$. Thus we
are left with the hierarchical perturbation equations
\begin{eqnarray}
y_{j}(x)&=&y_{0}(x)\int_{0}^{x}\frac{dx^{\prime }}{y_{0}(x^{\prime
})^{2}} \int_{0}^{x^{\prime }}\left( y_{0}y_{j-1}^{\prime
}\right.\\
&&\left.-y_{0}\sum_{k=1}^{j}E_{k}y_{j-k}\right) (x^{\prime \prime
})dx^{\prime \prime },  \label{eq:model1_yj}
\end{eqnarray}
where we have chosen $\alpha =\beta =0$ to satisfy the boundary
condition at $x=0$. Notice that $y_{j}$ depends on $E_{j}$ that is
determined by the boundary condition at $x=1$. For example, at
first order we obtain
\begin{equation}
y_{1}(x)=\frac{\sqrt{2}E_{1}x\cos (n\pi x)}{2n\pi }+\left( \frac{\sqrt{2}x}{2%
}-\frac{\sqrt{2}E_{1}}{2n^{2}\pi ^{2}}\right) \sin (n\pi x)
\label{eq:model1_y1}
\end{equation}
that does not satisfy the boundary condition at $x=1$ unless
$E_{1}=0$. Proceeding exactly in the same way we obtain the
function and energy perturbation
coefficients~(\ref{eq:model1_yj_Ej}) order by order, which clearly
shows that Eq.~(\ref{eq:model1_yj}) gives the correct answer.

The normalization factor calculated by perturbation theory
\begin{equation}
N(\lambda )=1-\frac{\lambda }{4}+\frac{\lambda ^{2}(n^{2}\pi ^{2}+12)}{%
96n^{2}\pi ^{2}}+\frac{\lambda ^{3}(n^{2}\pi ^{2}-12)}{384n^{2}\pi ^{2}}%
+\ldots   \label{eq:model1_N(lambda)}
\end{equation}
also agrees with the one derived from the exact solution.

\section{Partially solvable example \label{sec:nonsolvable}}

In order to compare our approach with Jaghoub's one~\cite{J06}
more closely, in what follows we treat one of that author's
examples:
\begin{equation}
\left( 1-\frac{3x^{2}}{5}\right) y^{\prime \prime
}(x)-\frac{6}{5}\left[ xy^{\prime }(x)-y(x)\right] +Ey(x)=0,
\label{eq:model3}
\end{equation}
where $y(0)=y(1)=0$ exactly as in the preceding example. Jaghoub
constructed this problem in order to have an exact solution for
the ground state~\cite {J06}
\begin{eqnarray}
y(x) &=&Ax(1-x^{2}),  \nonumber \\
E(n &=&1)=6  \label{eq:model3_y_E}
\end{eqnarray}
and treated only this trivial case by perturbation theory. Here,
on the other hand, we also consider the excited the states for
which there are no
exact solutions as far as we know. A straightforward calculation with $%
y_{0}(x)=\sin \left( n\pi x\right) $, $F=3x^{2}y^{\prime \prime
}/5+6(xy^{\prime }-y)/5-Ey$, and $F_{0}=-E_{0}y_{0}$ shows that
\begin{eqnarray}
E_{1} &=&-\frac{2n^{2}\pi ^{2}+15}{10},  \nonumber \\
y_{1}(x) &=&\frac{n\pi x(x^{2}-1)\cos (n\pi x)}{10}+\left( \frac{3x^{2}}{20}+%
\frac{1}{10}\right) \sin (n\pi x),  \nonumber \\
E_{2} &=&-\frac{3(8n^{4}\pi ^{4}+10n^{2}\pi ^{2}-15)}{1000n^{2}\pi
^{2}},
\nonumber \\
E_{3} &=&-\frac{248n^{6}\pi ^{6}+462n^{4}\pi ^{4}-1575n^{2}\pi ^{2}+1890}{%
35000n^{4}\pi ^{4}}.  \label{eq:model3_Ej_yj}
\end{eqnarray}
The energy coefficients agree with those shown numerically by
Jaghoub~\cite {J06} for $n=1$. In this case the normalization
factor is given by
\begin{equation}
N(\lambda )=\sqrt{2}\left[ 1-\frac{3\lambda
}{20}-\frac{3(29n^{4}\pi ^{4}+10n^{2}\pi ^{2}+30)\lambda
^{2}}{4000n^{4}\pi ^{4}}+\ldots \right]
\label{eq:model3_N(lambda)}
\end{equation}
We do not show the analytical expressions of $y_{2}(x)$ and
$y_{3}(x)$, as well as perturbation corrections of larger order,
because they are rather complicated.

\section{Conclusions \label{sec:conclusions}}

In this paper we have shown how to develop a useful perturbation
method from the Wronskian between the perturbed and unperturbed
solutions. The approach enables one to treat velocity--dependent
quantum--mechanical problems as well as local perturbation
potentials. Present approximation is more general than one
proposed earlier~\cite{J06} and enables us to discuss mathematical
aspects of the solutions that could be otherwise masked. We easily
derive the most general equation for the application of
perturbation theory and can clearly analyze all the contributions
to the approximate solutions and their behaviour regarding
normalization and boundary conditions.

We have tested our general equations on an exactly solvable model
showing that the approximate method yields exactly the same
results that one obtains from expansion of the exact solutions. We
have also treated a partially solvable problem. In both cases we
have derived perturbation corrections for all the states in terms
of the quantum number, thus generalizing Jaghoub's results for the
latter model~\cite{J06}.

Present perturbation method also applies to nonlinear models. We
may treat them exactly as the examples above, except that one has
to consider the normalization condition explicitly at every
perturbation order because the eigenvalues depend on it.

For simplicity we have chosen the unperturbed and perturbed
equations in such a way that we could solve all the integrals
analytically. In more difficult cases one should have to resort to
numerical integration. However, any problem that can be treated
exactly by Jaghoub's approach can also be treated exactly by
present one, probably in a more general way.

By means of an appropriate choice of the boundary conditions
present perturbation approach is suitable for the approximate
calculation of scattering phase shifts. In this field our method
may be an alternative to the logarithmic perturbation theory that
is commonly applied to local potentials~\cite{ACCLY92,ACC97}.

\begin{acknowledgement}
P.A. acknowledges support from Conacyt grant C01-40633/A-1
\end{acknowledgement}

\end{document}